\documentclass[notitlepage,11pt]{article}
\usepackage{epsfig}
\usepackage{fullpage}
\usepackage{subfigure}
\usepackage[sort&compress,numbers]{natbib}
\usepackage{wrapfig}

\begin{document}

\begin{center}
{\Large Superradiance as a Source of Collective Decoherence in Quantum Computers}

\vspace{0.2cm}

{\large D. D. Yavuz}

\vspace{0.2cm}

{\it Department of Physics, 1150 University Avenue,
University of Wisconsin at Madison, Madison, WI, 53706}

\end{center}

\vspace{1cm}

\noindent {\it \textbf{Abstract:}} We argue that superradiance (collective emission) due to radiative coupling of qubit states results in non-local noise, and thus introduces an error source that cannot be corrected using current models of fault-tolerant quantum computation. 

\newpage

Over the last decade, quantum computing and quantum information processing have emerged as very exciting fields of science due to the potential for solving exponentially large problems in polynomial time \cite{nielsen,bennett}. Initial enthusiasm for quantum computing was motivated in part by the polynomial-time prime factoring algorithm of Shor \cite{shor}. It has become increasingly evident that, in addition to factoring, quantum algorithms can be used for a broad class of problems, such as finding the eigenvalues and eigenvectors of large matrices \cite{jozsa,grover,abrams}. The fundamental building blocks of quantum computers are quantum bits, or qubits, which are used to store information. In principle,  qubits can be any quantum mechanical system that can be in two distinct states.  Computations are performed on qubits, just as operations are performed on classical bits, but qubit operations can exploit the extraordinary behavior of nature at the quantum scale. The principles of quantum computing have now been demonstrated using a variety of physical qubits including trapped ions \cite{cirac,monroe},  neutral atoms \cite{saffman,grangier}, semiconductor quantum dots \cite{loss}, superconducting Josephson junctions \cite{clarke}, and single photons \cite{knill}.  Currently many researchers around the world are working towards constructing a scalable quantum computer from these qubit building blocks.

One of the key achievements in the field has been the discovery of quantum error correction and fault-tolerant quantum computation \cite{shor2,steane}. In particular, the celebrated threshold theorem \cite{aharonov,terhal,aharonov2,preskill} has established confidence that if quantum gates are constructed with a fidelity better than a certain threshold, arbitrarily long quantum operations are, in principle, possible. Although the threshold theorem is a remarkable achievement, it has a key weakness: all formulations of the threshold theorem to date make certain assumptions regarding the properties of the noise that affects the quantum computer. In particular, the theorem works under the assumption that the noise must be both spatially and temporally local, affecting only a few qubits at a time. Physically, the locality of the noise is related to the interaction Hamiltonian that couples the qubits to the environment. Locality is guaranteed if the norm of the interaction Hamiltonian operator is bounded by a certain value. However, it has been pointed out that this assumption is not valid for certain models of environment-qubit coupling. In such cases, the threshold theorem becomes extremely sensitive to the high frequency spectrum of the environment operators \cite{preskill}. A number of authors have also expressed skepticism regarding the assumptions of the threshold theorem \cite{dyakonov,alicki}. 

In this letter, we argue that the well-known phenomenon of superradiance \cite{dicke,haroche} gives rise to noise that is not local.  By considering the radiative coupling of $L$ qubits to a common radiation bath in free space, we discuss that superradiance results in decoherence that scales with the square of the number of qubits. This decoherence mechanism is non-local in the sense that the collective emission simultaneously decoheres all the qubits in the computer. Because the noise source is not local, the introduced errors are outside the applicability of the threshold theorem. We also discuss that although the use of decoherence free subspaces (DFS's)  \cite{lidar1,lidar2,lidar3,whaley} will reduce the amount of non-local noise, they do not eliminate it completely. The reason for this is that any error in the preparation or the manipulation of the DFS will in general result in superradiance. As a result, residual non-local noise will remain, which will still be outside the applicability of the threshold theorem. Below we discuss radiatively coupled qubits in free space, so our results are particularly relevant for neutral atom- and trapped-ion-based quantum computation. However, our results will likely be applicable to other physical systems, since a source of collective noise can be found in most situations, for example, phonons for solid-state-based approaches. 

We begin our discussion by considering the interaction of $L$ two-level atoms with a continuum of radiation modes in the Schr\"{o}dinger picture. We will first assume the system to be in the superradiant regime, i.e., the system's physical size is small compared to the radiation wavelength. Although this may at first seem like a restrictive assumption, we will show below that it is not; the argument extends to the case where the spacing between the qubits is larger than the radiation wavelength. In the latter case, the essence of the argument is that, no matter how distant the qubits are, there are spatial modes of the radiation which will couple to all the qubits and induce collective emission. 

\vspace{0.2cm}

\noindent {\it \textbf{Formalism:}} We consider $L$ two-level atoms, each with levels $|g\rangle$ and $|e\rangle$. We will denote each individual atom with the index $j$. The levels of the $j$th atom will be labeled $|g\rangle^j$ and $|e\rangle^j$. We follow and extend the formalism of the Wigner-Weisskopf theory of spontaneous emission as described, for example, in Ref.~\cite{yamamoto}. A similar approach has recently been discussed in the analysis of reducing superradiance in the implementation of quantum algorithms \cite{yamamoto2}. We label the photon annihilation and creation operators for each mode $s$ by $\hat{a}_s$ and $\hat{a}_s^\dag$, respectively. The Hamiltonian for the combined atom-field system is:
\begin{eqnarray}
\hat{H}_{total}=\sum_{s}\hbar \nu_s \left( \hat{a}_s^\dag \hat{a}_s+\frac{1}{2} \right) + \sum_{j} \frac{1}{2}\hbar \omega_a \hat{\sigma}_z^j + \hat{H}_{int} \quad , 
\end{eqnarray}

\noindent where
\begin{eqnarray}
\hat{H}_{int} & = &- \sum_{j} \sum_s\hbar g_s \left( \hat{a}_s \hat{\sigma}_+^j+\hat{a}_s^\dag \hat{\sigma}_{-}^j \right) \quad , \nonumber \\
\hat{\sigma}_z^j & =& |e\rangle^j {^j}\langle e|-|g\rangle^j {^j}\langle g| \quad , \nonumber \\
\hat{\sigma}_+^j & =& |e\rangle^j {^j}\langle g| \quad , \nonumber \\
\hat{\sigma}_{-}^j & =& |g\rangle^j {^j}\langle e| \quad .
\end{eqnarray}

\noindent Here, $\hat{H}_{int}$ is the interaction Hamiltonian that determines the coupling between the atomic system and radiation modes. It is important to note that the Hamiltonian of Eq.~(2) uses the rotating wave approximation (RWA), i.e., only energy conserving terms are retained. The summation $\sum_{s}$ sums over all the relevant radiation modes. The energies of the atomic states $|g\rangle$ and $|e\rangle$ are taken to be $-\frac{1}{2} \hbar \omega_a$ and $\frac{1}{2} \hbar \omega_a$, respectively. We take the initial state of the atomic system to be an arbitrary (in general, entangled) superposition state and assume that initially each field mode $s$ is in vacuum state. The initial state of the combined atom-radiation field system can be written as:
\begin{eqnarray}
|\psi (t=0)\rangle = \sum_{q=0}^{2^L-1}c_{q,0} |q,0 \rangle \quad . 
\end{eqnarray}

\noindent Here, the 0 in state $|q,0 \rangle$ means that all modes $s$ have a zero photon number.  Following the superradiance literature, we define the following parameter for each atomic state $|q\rangle$:
\begin{eqnarray}
2 M_q \equiv  {\textnormal{\# of atoms in state $|e \rangle$}} -{\textnormal{\# of atoms in state $|g \rangle$}} \quad . 
\end{eqnarray}

\noindent With this definition, the energy of the atomic state $|q\rangle$ is $M_q \hbar \omega_a$. Working in the interaction picture, we expand the wavefunction as:
\begin{eqnarray}
|\psi(t) \rangle = \sum_{q=0}^{2^L-1} c_{q,0}(t) \exp{\left[ -i (M_q \omega_a)t \right]} |q,0 \rangle +\sum_s \sum_{q'=0}^{2^L-1} c_{q',s}(t) \exp{\left[ -i (M_{q'} \omega_a+\nu_s)t \right]} |q',1_s \rangle
\end{eqnarray}

\noindent Here, $|1_s \rangle$ represents the state of the radiation field in which the field mode $s$ has one photon while all the other modes are in vacuum state. With these definitions, we use the Hamiltonian of Eq.~(1) and derive the evolution equations for the probability amplitudes of Eq.~(5):
\begin{eqnarray}
\frac{dc_{q,0}}{dt}  & = & i \sum_s g_s \sum_{q' \in f(q)} c_{q',s}(t) \exp{\left[ -i (\nu_s-\omega_a)t \right]} \quad , \nonumber \\
\frac{dc_{q',s}}{dt}  & = & i  g_s \sum_{q'' \in g(q')}  c_{q'',0}(t) \exp{\left[ -i (\omega_a-\nu_s)t \right]} \quad .
\end{eqnarray}

\noindent Here, the index $q'$ runs through all the states that differ from $q$ by changing one atom from the excited level to the ground level. For example, if $q=|ee...e\rangle$, then $q'$ in  Eq.~(6) will run through a total of $L$ indices: $q'=|gee...e\rangle, |ege...e\rangle, ..., |eee...g\rangle $. The symbol $f(q)$ in the summation denotes this set of indices. Similarly, the index $q''$ runs through all the states that differ from $q'$ by changing one atom from the ground level to the excited level.  The symbol $g(q')$ in the summation denotes this set. 

To proceed, we follow the usual steps of Wigner-Weisskopf theory of spontaneous emission. The details of this derivation will be reported elsewhere. Briefly, we start by formally integrating the second line of Eq.~(6) and substituting the result into the first line. We then replace the summation $\sum_{s}$ with a frequency integral over the continuum of radiation modes. Performing the frequency integration under the usual assumptions, Eq.~(6) reduces to the following coupled differential equations for the probability amplitudes:
\begin{eqnarray}
\frac{dc_{q,0}}{dt}=-\left(\frac{\Gamma}{2}+i \delta \omega \right) \sum_{q' \in f(q)} \sum_{q'' \in g(q')} c_{q'',0}  \quad .
\end{eqnarray}

\noindent Here the quantities $\Gamma$ and $\delta \omega$ are, respectively,  the single atom decay rate and frequency shift (Lamb-shift) due to the coupling to the radiation continuum.
\vspace{0.2cm}

\noindent {\it \textbf{Dicke superradiance:}} Equation~(7) can be thought as the generalization of Dicke superradiance to an arbitrary initial superposition state.We note that for symmetric states, Eq.~(7) recovers the well-known results of Dicke superradiance \cite{dicke,haroche}. Assume that we start from an $L$ atom symmetric state where $\frac{L}{2}+M$ atoms are in the excited level and $\frac{L}{2}-M$ atoms are in the ground level. From Eq.~(7), each of the nonzero probability amplitudes in the symmetric state will evolve in an identical manner. We can therefore replace $c_{q'',0}(t)$ with $c_{q,0}(t)$ in Eq.~(7), which now reads:
\begin{eqnarray}
\frac{dc_{q,0}}{dt}=-\left(\frac{\Gamma}{2}+i \delta \omega \right) \sum_{q' \in f(q)} \sum_{q'' \in g(q')} c_{q,0} \quad .
\end{eqnarray}

\noindent The $\sum_{q'}$ summation will have $L/2+M$ terms and  $\sum_{q''}$ summation will have $L/2-M+1$ terms. Thus Eq.~(8) reduces to:
\begin{eqnarray}
\frac{dc_{q,0}}{dt}=-\left(\frac{\Gamma}{2}+i \delta \omega \right)\left(\frac{L}{2}+M\right) \left(\frac{L}{2}-M+1\right) c_{q,0} \quad . 
\end{eqnarray}

\noindent Hence, the new decay rate for the probability $\vert c_{q,0} \vert^2$ is:
\begin{eqnarray}
\left(\frac{L}{2}+M\right) \left(\frac{L}{2}-M+1\right)  \Gamma \quad , 
\end{eqnarray}

\noindent which is Dicke superradiance. For $M=\frac{L}{2}$ (all atoms in the excited level), the decay rate is $L \Gamma$. For $M=0$ (half of the atoms in the excited level, half of the atoms in the ground level), the decay rate is $\frac{L}{2}(\frac{L}{2}+1) \Gamma$. Finally, if $M=-\frac{L}{2}+1$ (only one atom is in the excited level), the decay rate is $L \Gamma$.

For an arbitrary initial state (not necessarily symmetric), the precise decay rate will depend on the initial values of $c_{q,0}(t=0)$ and can be calculated using the coupled equations of Eq.~(7). We note that quantum algorithms typically use a large portion of the state space, i.e., in general all $c_{q,0}(t=0)$ will be of comparable magnitude. Furthermore, there are exponentially more states with $M \sim 0$ than $M \sim L/2$. As a result, the vast majority of coefficients will decay as $ |c_{q,0}(t)|^2 \sim |c_{q,0}(t=0)|^2 \exp{\left[-(L^2/4) \Gamma  t\right]} $.

\vspace{0.2cm}

\noindent {\it \textbf{Non-local noise:}} We next discuss the non-local character of the noise due to superradiance. Physically, non-local character of the noise can be deduced from the fact that collective emission has contributions from and simultaneously decoheres all the qubits. As a result, superradiance noise cannot be assumed to affect only a few qubits at a time. In this section, we will make this argument more concrete by discussing the error on a single qubit during, for example, a quantum gate operation.  Consider a certain qubit at a specific location, $a$. We estimate the error on this specific qubit  by first calculating the reduced density matrix for this qubit. For this purpose, we write the initial wavefunction in a form where the levels of this qubit are explicit:
\begin{eqnarray}
|\psi (t=0)\rangle = \sum_{k=0}^{2^{L-1}-1} c_{k,0} |g\rangle^a \otimes |k,0 \rangle +  \sum_{k=0}^{2^{L-1}-1} d_{k,0} |e\rangle^a \otimes |k,0 \rangle \quad . 
\end{eqnarray}

\noindent Here, the index $k$ denotes the state of the remaining $L-1$ qubits other than the qubit at $a$. From the state vector of Eq.~(11), we form the density matrix for the global atomic system $\hat{\rho}=|\psi\rangle \langle \psi |$. We then obtain the reduced density matrix for the qubit $a$ by tracing over the coordinates of the remaining $L-1$ qubits, $\hat{\rho}_a= \textnormal{Tr}_k \left[ \hat{\rho} \right] $, which gives:
\begin{eqnarray}
\hat{\rho}_a=\sum_{k=0}^{2^{L-1}-1} \vert c_{k,0} \vert^2 |g\rangle^a {^a}\langle g| + \sum_{k=0}^{2^{L-1}-1} \vert d_{k,0} \vert^2 |e\rangle^a {^a}\langle e| +  \sum_{k=0}^{2^{L-1}-1} c_{k,0} d_{k,0}^* |g\rangle^a {^a}\langle e| + \sum_{k=0}^{2^{L-1}-1} c_{k,0}^* d_{k,0} |e\rangle^a {^a}\langle g| \quad .
\end{eqnarray}

\noindent In Eq.~(11), the states where the qubit $a$ is in the excited level $|e\rangle^a$ will have correlated emission from $L$ atoms. In contrast, if the qubit  $a$ is in the ground level $|g\rangle^a$, the system will have correlated emission from $L-1$ atoms. As a result, $ |c_{k,0}(t)|^2 \sim |c_{k,0}(t=0)|^2 \exp{\left[-((L-1)^2/4) \Gamma  t\right]} $ and $ |d_{k,0}(t)|^2 \sim |d_{k,0}(t=0)|^2 \exp{\left[-(L^2/4) \Gamma  t\right]} $. Using these expressions in Eq.~(12) gives a longitudinal and transverse decay rates of $L \Gamma /2$ and $L \Gamma /4$ for qubit $a$, respectively. For a sufficiently small gate time of $\Delta t$ we may approximate $\exp{\left(-L \Gamma \Delta t/2 \right)} \approx 1-L \Gamma  \Delta t/2$, which yields an error probability of $\epsilon=L \Gamma  \Delta t/2$ at the qubit $a$. This error is non-local since it scales with the number of qubits and the source of error simultaneously affects all the qubits.

It is important to note precisely how the formalism gives rise to non-local noise because this is in stark contrast with the current models of fault-tolerant quantum computation \cite{preskill}. In current models, having a constant error threshold requires the norm of the interaction Hamiltonian to be bounded by some finite value, i. e., $|| \hat{H}_{int} || < \infty$. The interaction Hamiltonian of Eq.~(1), which gives rise to superradiance, has an infinite norm and thus does not satisfy this requirement. 

\vspace{0.5cm}

\noindent {\it \textbf{Decoherence free subspaces:}} Over the last decade, the idea of using decoherence free subspaces (DFS's) has emerged as a promising way to reduce decoherence in quantum computers \cite{lidar1,lidar2}. The DFS's for the superradiance problem has been discussed in detail by a number of authors \cite{lidar3,whaley}. In this section, we argue that although the use of DFS's will reduce decoherence,  they do not eliminate non-local noise completely. The reason for this is that any error in the preparation or the manipulation of the DFS will couple the states to the larger Hilbert space and induce some collective emission. As a result, some residual non-local noise will remain, which will still be outside the applicability of the threshold theorem. 

Within our formalism, the existence of DFS's can be seen from the coupled equations of Eq.~(7). These form a linear set of equations and when written in a matrix form, will support a null-space for a specific set of initial values for the probability amplitudes, $c_{q,0}(t=0)$. It is well-known that for radiatively coupled $L$ qubits, a state of the DF subspace is a tensor product of the two-qubit singlet states \cite{lidar3}:
\begin{eqnarray}
|\psi\rangle _{DFS}=\left(\frac{1}{\sqrt{2} } \right)^{L/2} \otimes_{j=1}^{L/2} \left( |ge\rangle-|eg\rangle \right) \quad . 
\end{eqnarray}

\noindent With the initial values for the probability amplitudes given by Eq.~(13), it can be easily shown that the amplitudes will not change through the time evolution of Eq.~(7) (i.e., there is no decoherence). The state of Eq.~(13) cannot be prepared perfectly and at a specific point in the computation, one of the probability amplitudes may differ from its ideal value by a small amount. The dynamics of the system can then be investigated by numerically solving the coupled system of equations of Eq.~(7). We have performed this calculation for a different number of qubits in the computer and the results are displayed in Fig.~1. Here, we plot the initial rate of change of the probability amplitudes as the number of qubits in the system is varied. This rate can also be thought as the leakage rate of the error in $|\psi\rangle _{DFS}$ to the larger Hilbert space. The numerical results demonstrate a leakage rate of $(L^2/4) \Gamma$, which is the superradiant decay rate. Physically, this is because the perturbation to the DFS wavefunction results in mixing to the larger Hilbert space. The vast majority of the states in the larger Hilbert space are superradiant. 

\begin{figure}[tbh]
\vspace{-0.cm}
\begin{center}
\includegraphics[width=10cm]{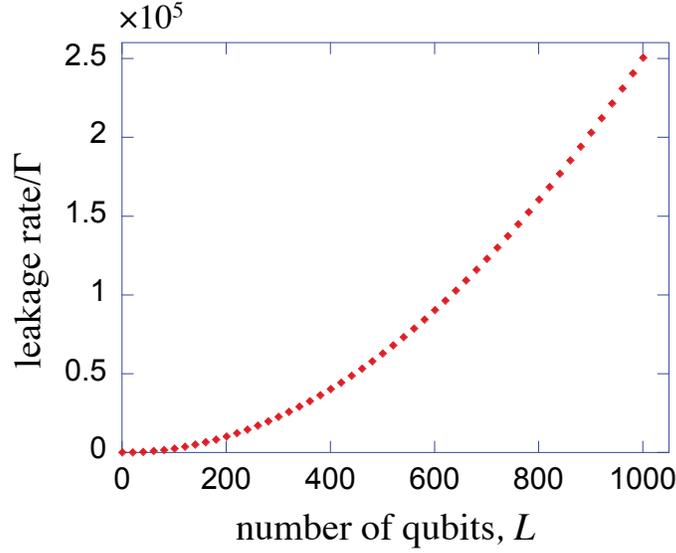}
\vspace{-0.4cm} \caption{\label{data} \small The leakage rate of the error in $|\psi\rangle _{DFS}$ as the number of qubits, $L$, is varied. The leakage rate equals $(L^2/4) \Gamma$ which is the superradiant decay rate. Since superradiance has contributions from all the qubits, this results in collective decoherence and therefore non-local noise even if the system is encoded in a DFS.  }
\end{center}
\vspace{-0.3cm}
\end{figure}

These results indicate that errors in the preparation or the manipulation of the DFS will result in superradiance. Since superradiance has contributions from all the qubits, this results in collective decoherence and therefore non-local noise. We therefore conclude that although working in a DFS will reduce decoherence, it will not eliminate collective decoherence (and therefore non-local noise) completely. 

\vspace{0.5cm}

\noindent {\it \textbf{Total decoherence during the computer run:}} These results suggest that collective emission produces noise that cannot be corrected using the current models of fault-tolerant quantum computation. In this section, we calculate the total decoherence of the system during the implementation of a quantum algorithm without any encoding in the DFS. As indicated by Eqs.~(7-10), the vast majority of the coherences of the full density matrix will decay as:
\begin{eqnarray}
\rho_{qq'}(t)=c_{q,0}(t)  c_{q',0}^*(t)\sim \rho_{qq'}(t=0) \exp{\left( -L^2 \Gamma t \right/4)} \quad . 
\end{eqnarray}

\noindent As above, let $\Delta t$ be the time it takes for a quantum gate operation. Also, let $R(L)$ be  the total number of gates that will be used in the computer run. With these definitions, the total computer run-time will be $R(L) \Delta t$. During this time the coherences will decay to:
\begin{eqnarray}
\rho_{qq'}(t=R(L) \Delta t) \sim  \rho_{qq'}(t=0)  \exp{\left( -L^2 \Gamma R(L) \Delta t /4 \right)} \quad . 
\end{eqnarray}

\noindent Note that Eq.~(15) can also be thought of as the success probability of the computer run:
\begin{eqnarray}
\mbox{success probability} \sim  \exp{\left( -L^2 \Gamma R(L) \Delta t /4 \right)} \quad . 
\end{eqnarray}

\noindent For a reasonably good success probability, we need to keep:
\begin{eqnarray}
L^2  R(L) \Gamma \Delta t /4<< 1 \quad \nonumber \\
\Rightarrow L^2  R(L) \Gamma /4 \omega_a<< 1 \quad . 
\end{eqnarray}

\noindent Here we have used the fact that the time required for a gate will be limited by the energy spacing between the two levels and, as a result, $\Delta t \sim 1/\omega_a$.

\vspace{0.2cm}

\noindent {\it \textbf{Qubits with a large spacing:}} In above, we have assumed the system to be fully in the superradiant regime where the total size of the computer is small compared to the wavelength of the radiation, $\lambda_a=2 \pi c/\omega_a$. In this section, we discuss that even when the system size becomes larger than the wavelength, the main results of the above argument remain the same. As shown in Fig.~2, we consider $L$ qubits in a three dimensional grid of size $w=L^{1/3}d$, where the spacing between the qubits may be much larger than the radiation wavelength, $ d >> \lambda_a$. Superradiance in large samples is known to be difficult to analyze in precise quantitative detail \cite{haroche}. However, a number of scaling results are well-known. As discussed in detail in Ref.~\cite{haroche} the modes of the electromagnetic radiation within the diffraction solid angle of the sample, $\sim (\lambda_a/w)^2$, interact with all the atoms and induce collective emission. To first order, superradiance in large samples can then be quantified by making the substitution $\Gamma \rightarrow \mu \Gamma $ and using the small-sample results. The parameter $\mu=\frac{3}{8 \pi^2}\frac{\lambda_a^2}{w^2}$ can physically be thought as the fraction of the vacuum modes that induce collective emission. Using this result, the collective emission rate is:
\begin{eqnarray}
\frac{L}{2}\left( \frac{L}{2}+1 \right) \mu \Gamma \sim \frac{3}{32 \pi^2} \frac{\lambda_a^2}{d^2} L^{4/3}  \Gamma \quad .
\end{eqnarray} 

\begin{figure}[tbh]
\vspace{0cm}
\begin{center}
\includegraphics[width=15cm]{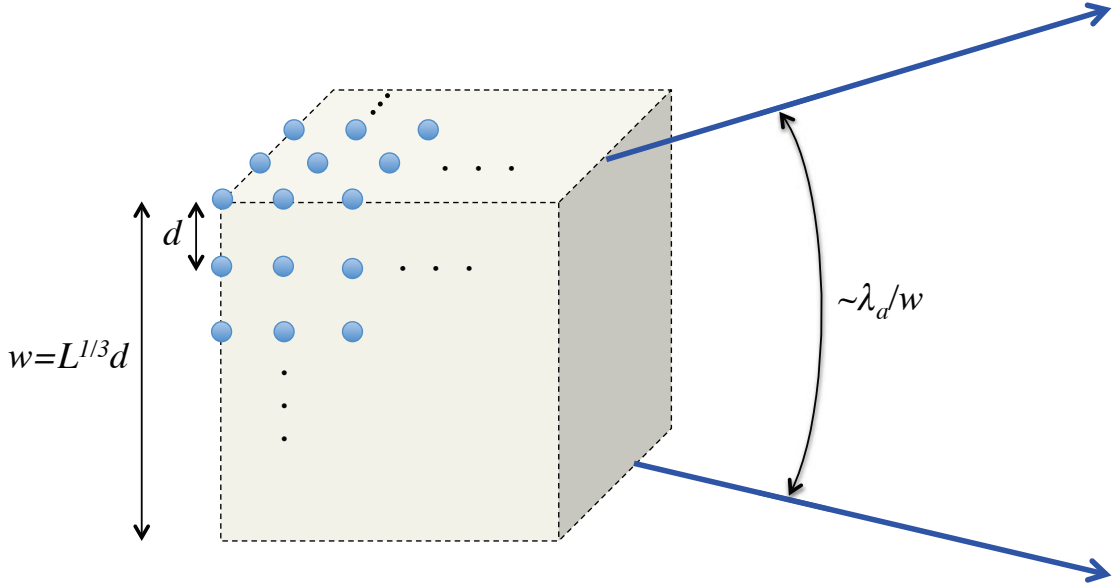}
\vspace{-0.4cm} \caption{\label{data} \small $L$ qubit quantum computer in a three dimensional geometry where the spacing between the qubits may be much larger than the radiation wavelength, $ d >> \lambda_a$. The modes of the radiation within the diffraction solid angle of the sample, $\sim (\lambda_a/w)^2$, interact with all the atoms and induce collective emission.  As a result, even for large samples,  there is a collective decoherence rate (and therefore non-local noise) which scales with the number of qubits. }
\end{center}
\vspace{-0cm}
\end{figure}

\noindent Equation~(18) indicates that even when the size of the computer is large compared to the radiation wavelength, there is a collective decoherence rate (and therefore non-local noise) which scales with the number of qubits. Compared to the small-sample case, there is one important difference: the rate of growth is $L^{4/3}$ instead of $L^2$. Similar to the above discussion, for a reasonable success probability of the computer run, we would need to keep
\begin{eqnarray}
\frac{3}{32 \pi^2} \frac{\lambda_a^2}{d^2} L^{4/3} R(L) \Gamma  \Delta t <<1 \quad ,
\end{eqnarray}

\noindent where $\Delta t$ is the time required for a quantum gate operation and $R(L)$ is the total number of gates. For the geometry of Fig.~(2), we can argue that each quantum gate between two qubits will at least require the speed of light propagation time between the two qubits, $\Delta t \sim d/c = ( 2 \pi/\omega_a) (d/\lambda_a)$. Hence Eq.~(19) can be reduced to:
\begin{eqnarray}
\frac{3}{16 \pi} \frac{\lambda_a}{d} L^{4/3} R(L)   \Gamma  \frac{1}{\omega_a} <<1 \quad .
\end{eqnarray}

\noindent We note that the total decoherence during the computer run as expressed in Eq.~(20) is an underestimate due to two key reasons: (i) While deriving Eq.~(20), we have only considered the nearest neighbor gates and used $\Delta t \sim d/c$ for the gate time. It is well-known that one can perform universal quantum computation using only nearest-neighbor gates; however there is in general an overhead. This overhead is not included in Eq.~(20). (ii) If an unrestricted architecture will be used, Eq.~(20) does not include the additional overhead (in the gate time) for performing gates between any two qubits within the computer. Both of these effects will increase the amount of decoherence during the computer run for the large-sample of atoms. A detailed study  of both of these effects is left for a future publication. 

\vspace{0.2cm}

\noindent {\it \textbf{Conclusions and acknowledgements:}} In conclusion, we have argued that superradiance produces an error rate on each qubit that scales as the number of qubits. Furthermore, the noise affects all the qubits simultaneously and is non-local. As a result, this type of noise cannot be corrected using the existing models of fault-tolerant quantum computation.  We believe our results give further importance to extending the current models of fault-tolerant quantum computation to non-local noise sources. We also note that there may be techniques to reduce the superradiance noise by modifying the density of states to suppress spontaneous emission, for example, by placing the qubits inside a high finesse cavity. Another alternative would be to use degenerate qubit states that are accessed by polarization selection rules. These will be among our future investigations. 

I would like to thank Mark Saffman and Thad Walker for introducing me to quantum computation and for many helpful discussions. Some of the ideas presented in this paper were developed while I was on sabbatical at the Electrical Engineering (EE) deparment at Bilkent University, Ankara/Turkey. I would like to thank \"{O}mer Morg\"{u}l and B\"{u}lent \"{O}zg\"{u}ler from Bilkent for helpful discussions. I also would like to thank other members of the EE department at Bilkent, particularly the chairman Orhan Ar{\i}kan, for their hospitality during my visit. Finally, I would like to thank two anonymous reviewers for helpful suggestions and comments.

\bibliographystyle{plainnat}

\end{document}